# Reducing Uncertainty in Simulation Estimates of the Surface Tension Through a Two-Scale Finite-Size Analysis: Thicker is Better

José L. Rivera[a,*] and Jack F. Douglas[b]

[a]Laboratorio de Modelamiento y Simulación Molecular, Universidad Michoacana de San Nicolás de Hidalgo, Morelia, Michoacán 58000, México

[b]Materials Science and Engineering Division, National Institute of Standards and Technology, Gaithersburg, Maryland 20899, USA

**Abstract**

Recent simulation studies of the surface tension $\gamma$, and other properties of thin free-standing films, have revealed unexpected finite size effects in which the variance of the properties vary monotonically with the in-plane width of the films, complicating the extrapolation of estimates of film properties to the thermodynamic limit. We carried out molecular dynamics simulations to determine the origin of this phenomenon, and to address the practical problem of developing a more reliable methodology for estimating $\gamma$ in the thermodynamic limit. We find that there are two distinct finite size effects that must be addressed in a finite size analysis of $\gamma$ in thin films. The first finite size scale is the in-plane width of the films and the second scale is the simulation cell size in the transverse direction. Increasing the first scale enhances fluctuations in $\gamma$, measured by the standard deviation of their distribution, while increasing the second reduces $\gamma$ fluctuations due to a corresponding increased 'freedom' of the film to fluctuate out of plane. We find that by using progressively large simulation cells in the transverse direction, while keeping the film width fixed to an extent in which the full bulk liquid zone is developed, allows us to obtain a smooth extrapolation to the thermodynamic limit, enabling a reduction of the $\gamma$ uncertainty to a magnitude on the order of 1 % for systems having a reasonable large size, i.e., O(1 μm).

*E-mail: jlrivera@umich.mx

**Introduction**

Longford et al.[1] recently studied the vapor/liquid equilibrium of Lennard-Jones atoms (Ar) and water molecules, using thin liquid films surrounded by coexisting vapor phases, and found unexpected finite-size effects for the variance of $\gamma$ as the width of the simulated film was varied. This result is important because it suggests that extrapolation of the film properties by increasing the film in-plane extent may not lead to reliable thermodynamic limit estimates of $\gamma$ and other film properties. For Lennard-Jones systems with liquid and vapor lengths, and dimensions in the transverse directions of between 15 $\sigma_{Ar-Ar}$ and 35 $\sigma_{Ar-Ar}$ ($\sigma_{Ar-Ar}$ being the diameter of a single Argon atom), temperatures between 85 K and 135 K, a short reduced-cutoff radius, $r_C^*$, of $\approx$ 3 $\sigma_{Ar-Ar}$, with long-range corrections to the properties, and a large timestep of 2 fs, the variance of the surface tension grows with the length of the layers in a linear fashion. Schmitz et al.[2,3] suggested that these unexpected size effects are due to a 'domain broadening' caused by density fluctuations in the bulk liquid. In a previous study, we found that bulk and some interfacial properties of narrow layers vary with the size of the simulation cell in the transverse directions, and that the average property values and their distribution narrow as the transverse area of the simulation cell is increased.[4] Here, we perform simulations aimed to understand these unexpected finite size effects, also observed of Langford et al.[1], in order to develop a more reliable approach for the precise estimation of the thermodynamic properties of thin films.

In this study, we examined two different finite-size effects on the variance of $\gamma$, where the variance is our focus because this quantity characterizes the distributions of our $\gamma$ estimations, which are normally distributed. First, by considering wider films composed of Lennard-Jones atoms, and longer $r_C^*$ to avoid the use of long-range potential corrections, we found a more complex behavior for the distribution of values of $\gamma$ than those reported by Longford et al.[1], who used a small $r_C^*$ with long-range corrections to the value of $\gamma$. We also found that the length of the simulation cell in the transverse direction makes the standard deviation of $\gamma$ decrease significantly as this scale increases, an effect apparently arising from the greater freedom of the interface to fluctuate out of plane when the transverse simulation scale is larger. There is evidently a two-scale finite size phenomenon arising in simulations, which is natural given the inherently anisotropic character of fluid films. Recognition of this two scale finite size effects allows us to develop a more reliable extrapolation method based on making the transverse film thickness larger while holding the in-plane films dimensions fixed, enabling the monotone extrapolation of film properties to thermodynamic limit.

**Methodology**

The properties of fluids are commonly calculated, in molecular dynamics simulations, as the mean value of data points that represent averages of blocks of timesteps calculated over up to millions of timesteps (which are on the scale of *fs*), with errors (standard deviation, variance, etc.) being calculated as a function of those average block values; however, this does not provide an idea of how the properties fluctuate during the dynamics of the systems. The use of block averages adds one more variable to the calculation process of properties, because this type of averaging can hide periodic phenomena related to natural phenomena such as the expansion and contraction processes in vapor/liquid equilibrium systems. The improper choose of the time-length of the blocks can lead to artifactual estimates for the calculated properties.

The vapor/liquid thermophysical and interfacial properties of the Lennard-Jones fluid was studied using molecular dynamics simulations at a reduced temperature $T^*$ of 0.72, which is in the vicinity of the triple point of the fluid. Direct simulations of interfaces are commonly used to study the vapor/liquid equilibria of pure and multicomponent systems, including polarizable systems, formed by thin layers of liquid surrounded by a vacuum or vapor phases in thermodynamic equilibrium.[5–10] The simulation cell consisted of a parallelepiped, with reduced dimensions of between 16 and 150.4 in the homogeneous directions (interfacial surface), and 48 in the inhomogeneous direction (which includes two interfaces), giving reduced interfacial areas between 256 and 22,620, and contained between 1040 and 336,346 Lennard-Jones atoms (Figure 1). Because there are two interfaces in our systems, we need to simulate vapor/liquid systems with enough vapor space to be in equilibrium with the liquid central layer. Systems with a ratio of 2 between the vapor and the liquid volumes are commonly employed,[11,12] so that liquid layers having a reduced thickness ≈ 16 requires a simulation cell having reduced length of 48 in the direction in the inhomogeneous direction. Performing the simulations near the triple point with this choice of cell anisotropy ensures that the capillary waves in the surface layer do not lead to the system becoming unstable. The initial systems consisted of solids with a face-centered cubic conformation, which were brought to the vapor/liquid equilibrium slowly, over a period of $10^6$ steps. The properties studied were calculated using additional simulations in an NVT ensemble (constant number of molecules, volume of the rigid simulation and temperature) for a period of $10^7$ steps, with a reduced timestep of 0.005. A Nosé[13] thermostat was used, implemented in the large-scale atomic/molecular massively parallel simulator (LAMMPS).[14]

Intermolecular interactions were calculated using the Lennard-Jones potential, which, in its reduced form, is:

$$U_{LJ}^* = 4\left[\left(\frac{1}{r_{ij}^*}\right)^{12} - \left(\frac{1}{r_{ij}^*}\right)^{6}\right], \qquad (1)$$

where $r_{ij}^*$ represents the reduced separation between Lennard-Jones atoms $i$ and $j$. The reduced Lennard-Jones potential was used with a $r_C^*$ of 7.5. By using this long $r_C^*$, all significant intermolecular interactions are taken into account, and the use of long-range corrections to calculate the total value of the properties is avoided.[15] The use of a short $r_C^*$, with long-range corrections, allows for reasonable predictions of the average vapor/liquid equilibrium properties, but the dynamics of the system are dependent on the $r_C^*$ employed. In this study, we used reduced variables and properties, which for fluids governed by the Lennard-Jones potential is an effective way of studying the corresponding states of Lennard-Jones fluids. The properties of fluids in laboratory units can be calculated from the reduced properties (indicated by the addition of an asterisk) using the $\sigma$ and $\varepsilon$ parameters of the Lennard-Jones potential associated with the specific fluid studied. Real lengths are calculated as $l = l^*\sigma$, time $t = t^*(m\sigma^2/\varepsilon)^{1/2}$, energy $U = U^*\varepsilon$, force $F = F^*\varepsilon/\sigma$, temperature $T = T^*\varepsilon/k_B$, density $\rho = \rho^*/\sigma^3$, pressure $P = P^*\varepsilon/\sigma^3$ and surface tension $\gamma = \gamma^*\varepsilon/\sigma^2$ and $k_B$ is Boltzmann's constant.

The pressure profiles were obtained through the calculation of pressure tensors in each slab, using Harasima pressure profiles,[16,17] implemented in LAMMPS,[18] in which the contributions to the profiles are distributed only in the two slabs that originated the interactions, and not evenly distributed through the slabs that lie between those that originated the interactions.[9,19] The normal pressure profiles obtained from this definition vary along the interface – a trend thought to have no physical sense, although it is not feasible to confirm these predicted profiles by measuring the components of the pressure in the interfacial region. Even so, the pressure profiles of Harasima have been used in several surface-tension studies because it does not matter whether the interaction forces are distributed uniformly in several slabs, or in only two; when it is integrated to obtain the profile of the surface tension, the same result is obtained.[10,20–23] As the contributions to the Harasima pressure profiles are located in the regions that originated the interactions, the density inhomogeneities at the interfaces are highlighted in the Harasima profiles. The Harasima normal pressure profiles are mechanically stable because the outermost regions show large negative pressures (attractive) holding together the whole liquid layer. Before the attractive zones, there are small peaks with positive pressure (repulsive), which, in comparison to the attractive ones, are smaller, and are not long and wide enough to burst the liquid layer. Here, the reduced surface tension $\gamma^*$ was calculated through its mechanical definition using the Harasima pressure profiles.[4,10] In this way, the whole system (two interfaces) is used to calculate de surface tension.

In order to obtain more realistic profiles, we allowed the system to move in the inhomogeneous direction, and calculated the density and pressures profiles every 100 steps. At the end, we averaged these by correcting the positions of the profiles according to their positions in the inhomogeneous direction of the center of the layer, which was calculated as the midpoint between the two positions of the Gibbs' dividing surface of each interface, $z_0^*$, with the thickness of the layer, $l_l^*$, corresponding to the separation between the positions of the two dividing surfaces. This allowed us to obtain more realistic profiles of the calculated properties. To obtain the positions of $z_0^*$, each of the density profiles, calculated every 100 steps, was adjusted to the hyperbolic tangent expression commonly used in vapor/liquid phase equilibrium studies:[11,19]

$$\rho^*(z) = \frac{1}{2}(\rho_l^* + \rho_v^*) - \frac{1}{2}(\rho_l^* - \rho_v^*)\tanh\left(\frac{z^* - z_0^*}{d^*}\right), \tag{2}$$

where $\rho_l^*$ and $\rho_v^*$ are the average bulk densities of the liquid and vapor phases, respectively, and $d^*$ is a measure of the thickness of the interface, and describes the length in the inhomogeneous direction, where the density changes from the bulk liquid to the bulk vapor phase.

**Results**

Molecular dynamics are commonly carried out in finite simulation cells, and it is important to consider those finite effects of the simulation cell on the property calculations.[24–26] Two finite-size variables that can be studied are the length of the layer and the interfacial area. Lengths of the layer below a critical thickness induce a layer break-up, while at the critical thickness, they produce metastable systems with no real bulk liquid phase.[4] Layers that are wide enough correspond to coexisting phases that are thermodynamically stable, and it is expected that further increments in the thickness will not change the interfacial and coexisting average properties. The size effects in the interfacial area are not intuitive due to the periodic boundary conditions employed, which try to reproduce the dynamics of systems with infinite and continuous interfacial areas; however, the periodic images create artificial cohesive forces at the interfaces, affecting certain thermophysical and interfacial properties, while the surface tension remains almost insensitive.[4] The cohesive forces are stronger in simulation cells having smaller interfacial areas, and result of the localization of the Lennard-Jones atoms, and thus enhancing the interatomic cohesive interactions, in the interfacial region. Due to the cutoff of the Lennard-Jones potential used, we performed the simulations using a reduced interfacial length, $b_t^*$, from 16 to 150.4, and thicknesses from the critical length to a reduced length of 16.8. Also, for the system with the smallest interfacial area used, we simulated systems with reduced lengths of up to 67.8.

The artificial cohesiveness forces induce a more structured behavior of the systems in the bulk phases and at the interfaces. Figure 2 shows the density profiles averaged over 100 timesteps for systems with $l_l^* = 16.8$ (far away from the critical length[4]), $T^* = 0.72$ and $b_t^* = 16$ and 150.4, which represent the smallest and largest surface areas studied. Even these profiles can be more well defined using more timesteps to average the profiles, the average over 100 timesteps allow us to understand the instantaneous density profile estimates. The system with a $b_t^*$ of 16 was more affected by the finite size area and produced larger fluctuations of the density not only in the bulk liquid, but also in the vapor phase and the interfaces. Density profiles averaged over $10^6$ timesteps or more produced smooth, and almost identical, density profiles, with very small difference in the average values in the bulk phases ($\Delta\rho_L^* = 0.01$), and equal to those previously reported from studies of phase equilibria, and obtained by Trokhymchuk and Alejandre using a $r_c^*$ of 5.5.[15] Even the average densities are almost equal for both $b_t^*$ values, the distributions of instantaneous bulk liquid densities are different. Using $b_t^* = 150.4$ produced a very narrow distribution of instantaneous bulk liquid densities when compared with the distribution at $b_t^* = 16$ (Figure 3). Similar behavior has been observed for distributions of the surface tension,[4] where the wide distributions are attributed to perturbations created by the induced cohesive forces, which are larger when small interfacial areas are used in the simulation cells. Comparing sections of $b_t^* = 16$ in the larger system with respect to the smaller system ($b_t^* = 16$), we can expect that the interfacial sections of size $b_t^* = 16$ in the larger system will have as much noise as the section of the smaller system, but these sections will interact collectively producing configurations with some sections having small surface tension, while in the other sections a relatively large surface tension is found due to out-of-plane expansions and contractions of the film. These fluctuations narrow the distribution of values of the surface tension and liquid density in larger systems.

In a similar way, the average $\gamma$ is also insensitive to the value of $b_t^*$ employed in the simulation, as previously reported,[4] but also it is insensitive to the length of the layer, as shown in Figure 4. At constant $b_t^*$, the average values of $\gamma$ remain insensitive as $l_l^*$ grows, while the corresponding standard deviation grows. The standard deviation values of $\gamma$ did not reduce its magnitude when they were calculated over longer periods of simulation, probably indicating that smaller standard deviations are only produced when standard deviations are calculated using average values of $\gamma$ of blocks of time, probably hiding periodic phenomena. When comparing the average values of the two sets at different $b_t^*$, for each $l_l^*$, we found very small differences that were less than the size of the symbols. On the other hand, we found a similar behavior between the distributions of the bulk liquid density and the distributions of the surface tension, in which the distributions of $\gamma$ narrowed as $b_t^*$ increased, which manifested as lower standard deviations.

We studied the origins of the insensitivity of the behavior of the average value of $\gamma^*$ with the size of $b_t^*$ at constant $l_l^*$ through analyzing the components of the average pressure profiles. In Figure 5, we plotted the normal ($P_N^*$), transverse ($P_T^*$), and difference between the average normal and average tangential pressure profiles ($\Delta P^* = P_N^* - P_T^*$) for systems with $l_l^* = 16.8$ and $b_t^* = \{16, 140.8, 150.4\}$. The $P_T^*$ profiles show the regular behavior of near-zero values in the bulk zones and large negative peaks at the interfaces, which are the result of the inhomogeneous density at the interface. At the interfacial zones, the average $P_T^*$ profiles show clear differences between the system simulated using $b_t^* = 16$ and the systems simulated at larger $b_t^*$ sizes. In the reduced units, the minimum in the negative peak for the system simulated using $b_t^* = 16$ is ≈ 0.09 deeper than the systems simulated using larger $b_t^*$ sizes, while at the outermost part of the interface, the profile using $b_t^* = 16$ shows slightly lower values. The systems simulated at $b_t^*$ sizes of 140.8 and 150.4 are identical, and reflect the visual disappearance of the size effects on the average $P_T^*$ profiles using large $b_t^*$ sizes.

The average $P_N^*$ profiles also show the regular behavior of the Harasima pressure profiles, which are not flat at the interface. They depict a more intuitive picture, wherein the outermost monolayer of atoms is strongly attracted by the bulk liquid, originating the outermost negative peak (cohesive) on the average $P_N^*$ profiles. The strongly attracted outermost monolayers crunch a monolayer of atoms located between them and the bulk liquid, which traduces in the positive peaks (repulsive) located in the reduced positions of $\pm 7.5$ ($b_t^* = 16$) and $\pm 7.7$ ($b_t^* = \{140.8, 150.4\}$). The average $P_N^*$ profiles also show differences depending on the size of $b_t^*$ employed during the simulation. The maxima and minima in the peaks for the system simulated using $b_t^* = 16$ are larger by ≈ 0.03 (positive peaks) and ≈ -0.06 (negative peaks) reduced pressure units than the systems using larger $b_t^*$ sizes. Close to the outermost part of the interface, the average $P_N^*$ profiles, using $b_t^* = 16$, also show lower values than the profiles using larger $b_t^*$ values, which can affect the processes of vaporization/condensation, which were beyond the scope of this study. The integration of the average Harasima $P_N^*$ profiles results in a net contribution identical to the integration of a flat $P_N^*$ profile, which some authors have assumed has a physical sense (Figure 6a). The slope of the net contribution is not zero because the average $P_N^*$ in the bulk zones is not zero, instead it corresponds to the saturation pressure.

The average $\Delta P^*$ profiles also show a regular behavior, with large positive peaks located at the interfaces and small negative peaks located in the outermost parts of the interface. The positions of the large positive peaks are identical for small and large sizes of $b_t^*$. The maxima and minima in the peaks for the system simulated using $b_t^* = 16$ are larger by ≈ 0.07 (positive) and ≈ - 0.02 (negative) reduced pressure units than the systems using large $b_t^*$ sizes. The negative minimum for the system simulated using $b_t^* = 16$ is located at 0.5 reduced length units off the position of the systems using larger $b_t^*$ sizes. When we integrated these profiles, we obtained the $\gamma^*$ profiles (Figure 6b). Even the $\Delta P^*$ profiles show small differences depending on the $b_t^*$ value employed, with the $\gamma^*$ profiles being identical. Therefore, we can attribute the insensitivity of $\gamma^*$ to $b_t^*$ as the result of a balance between the positive contributions of $P_N^*$ (positive peak) and $-P_T^*$ and the negative contribution of $P_N^*$ (negative peak). For the simulated system using $b_t^* = 16$, the area of the positive peak in the average $\Delta P^*$ profiles indicate a larger surface tension contribution than the peaks obtained using larger $b_t^*$ sizes, but the outermost negative contributions are also larger for $b_t^* = 16$, which match the larger positive contributions, resulting in an annulment of the additional contributions.

The study of the pressure profiles at constant $b_t^*$ allowed us to examine the insensitivity of the behavior of the average value of $\gamma^*$ with the size of $l_l^*$ (Figure 7). The plotted $l_l^*$ values corresponds to $l_l^*$ of 8.0, 12.6 and 16.8 at a constant $b_t^*$ of 120. The systems simulated using short $l_l^*$ sizes (8.0) did not develop a real bulk liquid phase in which the values of the pressure profiles reached the same values as the bulk vapor phase. This system seems to be composed of two joined interfaces. As the size of $l_l^*$ grows, the bulk liquid zone develops, and the pressures in all profiles reach values similar to those in the bulk vapor phases. The magnitudes for all peaks for all $l_l^*$ are identical in the three pressure profiles ($P_N^*$, $P_T^*$ and $\Delta P^*$). When the $\Delta P^*$ profiles are integrated to obtain the $\gamma^*$ profiles of the two interfaces (Figure 8), all profiles with different $l_l^*$ produce the same average magnitude of $\gamma^*$. The profiles of $\gamma^*$ for systems using larger $l_l^*$ show a central flat region, which correspond to the bulk liquid zone, which reduces until it disappears at the shortest size of $l_l^*$ reported. Even the $\Delta P^*$ profiles using short sizes for $l_l^*$ do not reach equilibrium values when they are integrated to produce the $\gamma^*$ profiles; instead, they produce the same average $\gamma^*$ calculated using larger $l_l^*$ sizes (Figure 8). As $l_l^*$ decreases, the interfaces seem to join (Figure 7), producing a central zone characterized by large normal (positive) and tangential (negative) pressures. As the interfaces join, the pressures of the interfaces, which give rise to the surface tension, do not cancel out or disappear, rather they accumulate in the short central zone, ultimately producing the same average value for the surface tension due to the symmetry of the two interfaces in the $\Delta P^*$ profile.

An insensitivity of the average $\gamma^*$ to $b_t^*$ is expected in capillary wave theory,[27] in which the interfacial thickness, $\Delta$, can be decoupled into an intrinsic contribution, $\Delta_0$, and a logarithmic contribution. The interfacial thickness depends on the size of the simulation cell in the transverse direction (which directly correlates with the size of the capillary waves that can form at the interface), while the average $\gamma$ remains constant. Larger simulation cells induce wider capillary waves, making the interfacial thickness also wider (at very large $b_t^*$ values, probably unrestricted). In laboratory units:

$$\Delta^2 = \Delta_0^2 + \frac{k_B T}{2\pi\gamma} \ln\left(\frac{b_t}{B_0}\right), \tag{3}$$

where $B_0$ is a characteristic size of the simulation cell in the transverse direction. At very short $b_t$ sizes (close to $B_0$), the interfacial thickness is equal to their intrinsic value. A study on water simulations have employed Equation 3 to predict the surface tension at a specific thickness of the layer,[27] but no comparisons have been made at different thicknesses of the layer, and none have explored the limits of Equation 3 at very narrow thicknesses. Figure 9 shows $(\Delta^*)^2$ as a function of $ln(b_t^*)$ for the systems with $l_l^*$ of 8.0 and 16.8. Linear regressions of the two sets of data produce almost parallel lines, which also reflect the insensitive behavior of $\gamma^*$ not only with $b_t^*$, but also with $l_l^*$.

The standard deviation of $\gamma^*$, $\sigma_\gamma^*$, is plotted as a function of $l_l^*$ in Figure 10 for the simulated systems using sizes of $b_t^*$ of 16, 40, 120 and 150.4. We calculated the profiles of $\sigma_\gamma^*$ for each $b_t^*$ from the first stable layers at its critical length,[4] with thicknesses corresponding to a few monolayers, up to a reduced value of $l_l^* = 16.8$. The profiles show increasingly smaller values of $\sigma_\gamma^*$ as $b_t^*$ increases. The inner graph of Figure 10 shows additional data results for the smallest size of $b_t^*$ used (16) for thin layers with $l_l^*$ up to 67.8; the profiles of $\sigma_\gamma^*$ seem to follow a power function behavior with $l_l^*$. The profiles of the simulated systems ($l_l^* \leq 16.8$) using larger values of $b_t^*$ seem to follow a near-linear behavior with $l_l^*$ as $b_t^*$ grows. The increasingly linear behavior of the profiles with $b_t^*$ can be modeled through a power function of the form:

$$\sigma_\gamma^* = c_1 + c_2 (l_l^*)^{c_3}, \tag{4}$$

which produces values for the $c_3$ power exponent of 0.43, 0.44 and 0.49 for the systems with $b_t^*$ of 16, 40 and 120, respectively. For the thin layer using $b_t^* = 150.4$, a linear behavior is the best fit. For the systems simulated using the smallest size of $b_t^*$ (16), the parameters of Equation 4 fitted with thinner layers ($l_l^* \leq 16.8$), produced values close to those of the simulation results for the wider layers ($16.8 \leq l_l^* \leq 67.8$), and are also shown in the inner plot of Figure 10, which reflects the consistency of the obtained parameters using data from the smaller $l_l^*$ systems. We note that our $\sigma_\gamma^*$ with $l_l^*$ estimates do not follow the root square dependence with system size. This unexpected behavior is because the interfacial forces vary with $l_l^*$ in thin-layer systems, as can be seen in Figure 10.

Yoon et al.[28] have proposed that only a few layers of the interface should be used to calculate the surface tension to avoid the size effects of the bulk phases (length of the layer). Yoon et al.[28] employed this methodology to calculate the surface tension of water using a region of $\approx 5$ Å (which is less than the diameter of 2 water monolayers) in a small system composed of 804 water molecules. The limits of this region corresponded to 0.01 and 0.90 g/cm³, which are asymmetrical in the density profile of water, and do not correspond to the values of the so-called "10-90" thicknesses, which corresponds to limiting points in the density profile at 10% and 90% of $\rho_l^*$. The final result of Yoon et al.[28] calculations did not produce a consistent tendency in the surface tension of water as a function of temperature. We employed this strategy to calculate the surface tension of the Lennard-Jones fluid by eliminating regions of the bulk phases, and the results are summarized in Table I. If we eliminate parts of the bulk liquid from the integral to compute $\gamma^*$, $\gamma^*$ and $\sigma_\gamma^*$ decreased, reaching a minimum when we eliminated a region that represents $\approx 70\%$ of the layer thickness, beyond this point the behavior is ill-defined. The decrements in $\gamma^*$ were then less pronounced than the decrements in $\sigma_\gamma^*$, but still the maximum reduction of $\sigma_\gamma^*$ was half its original value, produces considerable changes in $\gamma^*$ ($\approx -20\%$). If we eliminate parts of the bulk vapor from the integral to compute $\gamma^*$, the resulting changes in $\gamma^*$ and $\sigma_\gamma^*$ are minimal even when we eliminate a region limited by a point located 0.5 reduced units away from the Gibbs' dividing surfaces. Eliminating at the same time parts of both bulk phases as proposed by Yoon et al.[28] does not produce a zero change in $\gamma^*$ unless you eliminate a very small region in the bulk liquid, but those changes will not reduce considerably $\sigma_\gamma^*$. These results indicate that the integration over the bulk vapor region does not contribute significantly to the average value of $\gamma^*$, neither $\sigma_\gamma^*$; the main contribution comes from the region between the interface and a point located between 6.5 and 8.5 reduced units (monolayers) away from the interface, in this zone the liquid density and pressures are uniform and the pressures are equal (bulk). For these narrow layers (nanometer scales), the distributions of values of $\gamma^*$ in fact depend on the size of the bulk liquid, no from both bulk phases, because the contractions/expansions that occur at the interface have to be the result of expansions/contractions in the bulk liquid phase, which ultimately indicates that interfaces of narrow layers do not expand/contract independently of the size of the layer.

We then studied the possibility that the large ratio between the tangential length and the normal length of the simulation cells used in this work can create an artefact on the calculated properties and standard deviations. For the largest system studied, which used a $b_t^* = 150.4$ and $l_l^* = 16.8$, having originally a ratio of 3.12, we simulated systems with dimensions of 96, 192 and 288 reduced length units in the normal direction, which had ratios of 1.56, 0.78 and 0.52, respectively (Supplementary Information, Figure S1). This additional space was spontaneously filled by vapor, and more molecules were added to maintain the $l_l^* = 16.8$. After this procedure, we did not observe any change in the average density profiles or the average pressure profiles (See Figures S2 and S3 of Supplementary Information). The average values of $\gamma^*$ and $\sigma_\gamma^*$ are thus not affected by the aspect ratio between the cell dimensions in the tangential and normal directions, having a ratio between the volumes of the vapor and liquid phase is enough. These observations are in concordance with our previous assessment of the methodology of bulk phase elimination proposed by Yoon et al.[28].

Longford et al.[1] developed an expression showing the dependency of the variance of $\gamma^*$, $Var_\gamma^*$, which is the square of $\sigma_\gamma^*$, with respect to a variable that combines the two finite size effects, $l_l^*$ and the interfacial area of the simulation cell, $A^* = (b_t^*)^2$,

$$Var_\gamma^* = c_4 \frac{l_l^*}{A^*} - c_5 \frac{r_c^*}{A^*}. \qquad (5)$$

where $c_4$ and $c_5$ are constants. Figure 11a shows the calculated values of $Var_\gamma^*$ as a function of $l_i^*/A^*$. Plotting $Var_\gamma^*$ vs. $l_i^*/A^*$ produces linear sets of data, almost parallel, and with different intercepts to *y*-axis depending on $A^*$, which shows significant contributions of the second term of Expression 5. The slight visual differences in the profile for $b_t^*=16$ are probably the result of the non-quadratic behavior of $\sigma_\gamma^*$ with $l_i^*$ (Equation 4). The intercepts with *y*-axis show a linear behavior when they are plotted against the variable $1/A^*$ (Figure 11b) in agreement with Expression 5. The simulations of Longford *et al.*[1] using a $r_C^* \sim 3$ (10 Å) show a linear behavior with almost-zero contributions of the second term of Expression 5. The different behavior is clearly due to the large $r_C^*$ used in this work (7.5), and predicted by the expression developed by Longford *et al.*[1] The surface tension results of Longford *et al.*[1] included long-range corrections to the surface tension, those corrections are based on the density profiles, but for systems under the same coexisting bulk densities and same $r_C^*$, those contributions should also be the same, therefore the observed null contributions of the second term in the work of Longford *et al.*[1] are expected, because the dynamics of the systems, and therefore the distributions of $\gamma^*$ are developed basically at the $r_C^*$ of 3.

A plot of $\sigma_\gamma^*$ as a function of $(b_t^*)^{-1}$ at constant $l_i^*$ allowed us to predict the behavior of $\sigma_\gamma^*$ at long $b_t^*$ values. Figure 12 shows $\sigma_\gamma^*$ as a function of $b_t^*$ for two thin layers with $l_i^*$ of 8.0 and 16.8. The two sets of data show an asymptotic behavior of $\sigma_\gamma^*$ with $b_t^*$, in agreement with the expression developed by Longford *et al.*[1], which can be modeled as an inversely asymptotic function of the form:

$$\sigma_\gamma^* = \frac{c_6}{b_t^*}, \qquad (6)$$

which produces values for the $c_6$ constant for the system with $l_i^* = 8.0$, of 5.26 (correlation coefficient of 0.9968). For the system with $l_i^* = 16.8$, the constant is 11.06 (0.9999). The lower agreement for the data using $l_i^* = 8.0$ probably is the result of the small thickness of the layers, which are closer to the critical thickness. In the hypothetical case that we wanted to measure $\gamma^*$ at $T^* = 0.72$, using a very large simulation cell in the tangential direction to reduce the fluctuations of $\gamma^*$ to less than 1% of its total value, we could not do it with a thin layer with $l_i^* = 8.0$ as this value is lower than the expected critical thickness free of the size effects,[4] and with a layer with $l_i^* = 16.8$, we would need a simulation cell with $b_t^* \approx 1050$, which corresponds to a simulation cell with tangential direction dimensions at the micrometer scale (for argon or methane).

The $\sigma_\gamma^*$ profiles not only become more linear as $b_t^*$ increases, but the slope of the curves decreases (from positive values) and becomes flatter as $b_t^*$ increases. In Figure 13, we show the calculated slope of $\sigma_\gamma^*$ with $b_t^*$ for the systems with $l_i^* = 16.8$. Comparing the systems using $b_t^*$ of 40 and 150.4, the results indicate that increases in the fluctuations as large as the value of $\gamma^*$ are obtained when we increase the thickness of the layer by $\approx 80$ reduced units of length at $b_t^* = 40$, and $\approx 380$ reduced units of $l_i^*$ at $b_t^* = 150.4$. The set of data also seems to follow an inversely asymptotic behavior, in agreement with Expression 5, and the data was fitted to the corresponding expression:

$$\frac{\partial \sigma_\gamma^*}{\partial l_i^*}(l_i^* = 16.8) = \frac{c_7}{b_t^*}, \qquad (7)$$

which produces values for the fitting coefficient $c_7$ of 4.14 x 10$^{-1}$ using the 4 largest $b_t^*$ values (correlation coefficient of 0.9980). The results for the smaller values of $b_t^*$ deviates positively from the predicted behavior. This model indicates that there is a large value for $b_t^*$ at which $\sigma_\gamma^*$ does not grow or grows very slowly as $l_i^*$ increases.

**Conclusions**

Extrapolations of data obtained from molecular dynamics simulations of the vapor/liquid equilibrium of Lennard-Jones atoms at a temperature close to the triple point, have shown that simulation cells with lengths at the microscale are needed to minimize the fluctuations (standard deviations $\leq 1\%$) of the surface tension, and probably other physicochemical properties, present when narrow liquid layers are simulated (composed of $\approx 17$ monolayers). Narrower liquid layers, composed of ~8 monolayers, will never reach such small standard deviations. Probably, liquid layers wider than 17 monolayers will need simulation cells with tangential lengths shorter than the microscale.

The behavior of the distributions of the surface tension, in terms of its standard deviation, shows an opposite dependence on two finite-size parameters: the length of the simulation cell in the tangential direction reduces the magnitude of the standard deviation as longer lengths are employed; and the thickness of the layer makes the standard deviation of the surface tension grow as wider layers are simulated. The dependence with the length of the layer disappears when larger lengths in the tangential direction are employed. Within the range of the systems simulated, we did not find asymptotic points for the standard deviation of the surface tension, but we did find a clear trend that can be modeled by asymptotic functions. The variance of the surface tension shows a previously observed linear behavior with respect to a combining variable that includes the two size effects, with large intercepts to the *y*-axis that depend linearly on the interfacial area of the simulation cell. The distribution of values of the surface tension using the large $r_C^*$ employed in this work cannot be reproduced using small $r_C^*$ and post-simulation long-range corrections.

The effect of the length of the simulation cell in the transverse direction is due to the induction of cohesive forces at the interfaces in both directions, which are the result of the employment of periodic conditions, which have stronger effects when short interfacial areas are employed. When very narrow liquid layers are simulated (near the critical length), the interfaces of the system interact with each other, probably coordinating the interfacial behavior of each other. As wider lengths of the liquid layer are employed, the interfaces become freer, and there is no coordination between these, making them free to develop their own dynamics.

The size effect due to the finite size of the simulation cell in the interfacial area is clearly an artefact of the methodology employed to simulate phase equilibria under periodic conditions in the interfacial surface, but for systems confined to these interfacial-length scales, probably this size effect has a physical meaning and future work should elucidate its influence. The size effect due to the thickness of the layer is only important for narrow layers in the scale of the nanometer, beyond that its influence decreases until they disappear at larger thicknesses layers. In the future, we also plan to assess the effects of additives and multicomponent equilibria to study the dependence of finite size effects on the distributions of their interfacial properties.

**Conflicts of interest**

There are no conflicts of interest to declare.

**Acknowledgments**

We thank Francis W. Starr from Wesleyan University for his valuable comments to the paper. We thank CONACYT (México) for an infrastructure fellowship and the Universidad Michoacana de San Nicolás de Hidalgo for research funds.

**Table I.** Simulation results of the surface tension and its standard deviation for the system of Lennard-Jones atoms using a $b_t^* = 150.4$, $l_i^* = 16.8$, and $T^* = 0.72$. The integration to obtain the surface tension was carried eliminating the indicated regions of the bulk liquid, or the bulk vapor phases. The case when the whole system is integrated is presented for comparison. The whole system covers a region in reduced units of (-24 − 24), the bulk liquid is located at the center, and the Gibbs' dividing surfaces located at positions in reduced units of -8.4 and 8.4.

| Eliminated region (positions in reduced units) | Eliminated thickness (reduced units) | $\gamma^*$ (% change) | $\sigma_\gamma^*$ (% change) |
|---|---|---|---|
|  | 0 | 1.051 (0) | 7.345 x 10⁻² (0) |
| Liquid, (-1 − 1) | 2 | 1.050 (-0.09) | 6.819 x 10⁻² (-7.08) |
| Liquid, (-2 − 2) | 4 | 1.048 (-0.28) | 6.241 x 10⁻² (-15.03) |
| Liquid, (-3 − 3) | 6 | 1.041 (-0.95) | 5.602 x 10⁻² (-23.73) |
| Liquid, (-4 − 4) | 8 | 1.023 (-2.66) | 4.905 x 10⁻² (-33.22) |
| Liquid, (-5 − 5) | 10 | 0.974 (-7.32) | 4.179 x 10⁻² (-43.10) |
| Liquid, (-6 − 6) | 12 | 0.850 (-19.12) | 3.777 x 10⁻² (-48.57) |
| Liquid, (-7 − 7) | 14 | 0.566 (-46.14) | 4.464 x 10⁻² (-39.22) |
| Liquid, (-8 − 8) | 16 | 0.172 (-83.63) | 3.726 x 10⁻² (-49.27) |
|  |  |  |  |
| Vapor, (-24 − -11) ∪ (11 − 24) | 26 | 1.052 (0.09) | 7.348 x 10⁻² (0.04) |
| Vapor, (-24 − -10) ∪ (10 − 24) | 28 | 1.059 (0.76) | 7.361 x 10⁻² (0.21) |
| Vapor, (-24 − -9) ∪ (9 − 24) | 30 | 1.059 (0.76) | 7.407 x 10⁻² (0.84) |

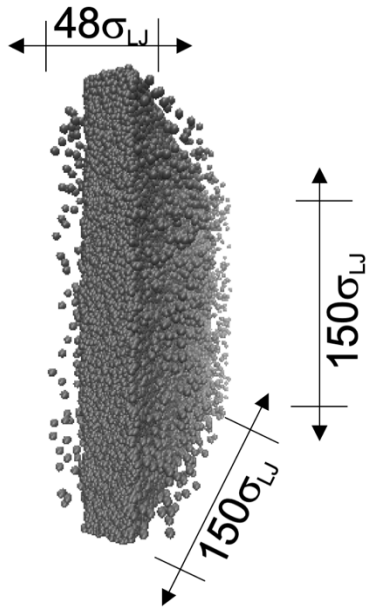

**Figure 1**. Orthogonal projection of a snapshot of the central cell used in the molecular dynamics simulations. The marked dimensions correspond to the reduced lengths for the simulation cell with the widest interfacial area studied. Each sphere represents a Lennard-Jones atom at a reduced temperature of 0.72.

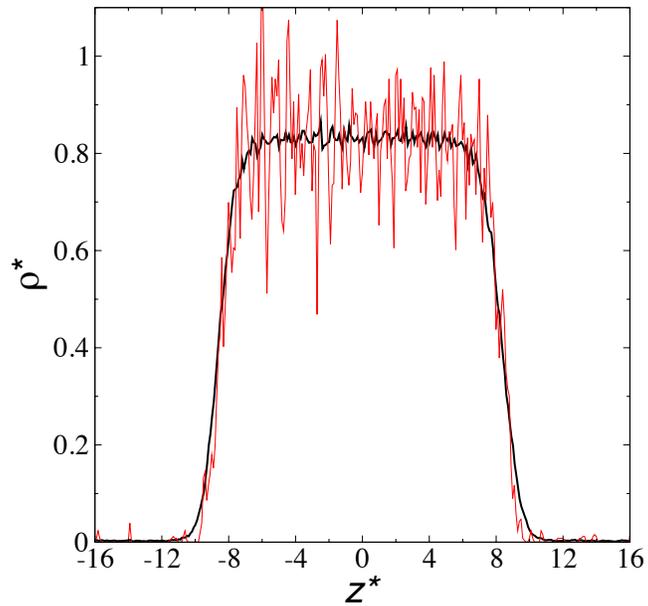

**Figure 2**. Reduced density profiles as a function of the position in the inhomogeneous direction of the simulation cell. The profiles represent average profiles of 100 timesteps. The systems are composed of Lennard-Jones atoms at a reduced temperature of 0.72 and reduced thicknesses of ≈ 16.8. Red and black lines correspond to simulated systems using reduced lengths of the simulation box in the tangential direction of 16 and 150.4, respectively.

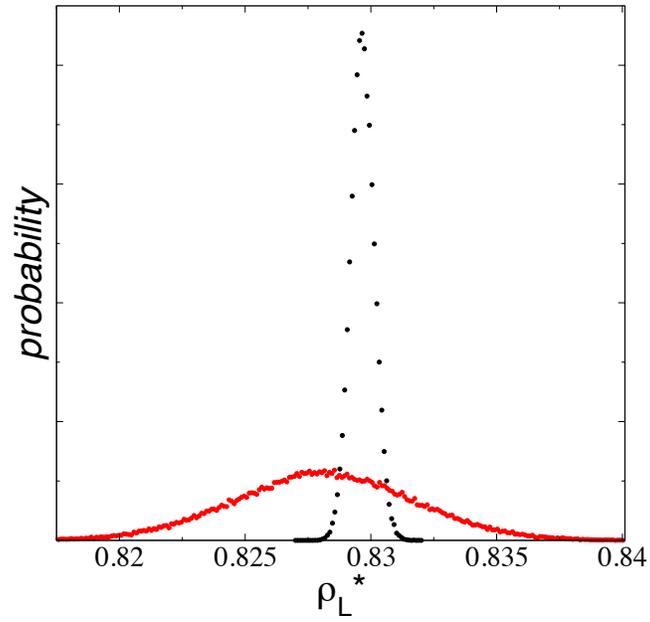

**Figure 3.** Histograms of the liquid densities averaged over 100 timesteps. The systems are composed of Lennard-Jones atoms at a reduced temperature of 0.72 and reduced layer thicknesses of ≈ 16.8. Red and black profiles correspond to simulated systems using reduced lengths of the simulation box in the tangential direction of 16 and 150.4, respectively.

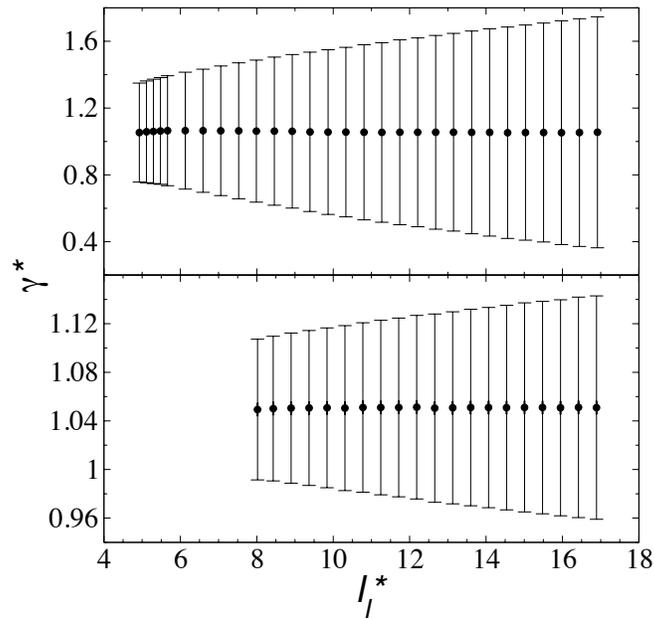

**Figure 4.** Surface tension as a function of the length of the layer for simulations cells with reduced tangential lengths of 16 (top) and 120 (bottom). The systems are composed of Lennard-Jones atoms at a reduced temperature of 0.72. The shortest lengths correspond to layers at their critical lengths.[4] The error bars correspond to their standard deviations.

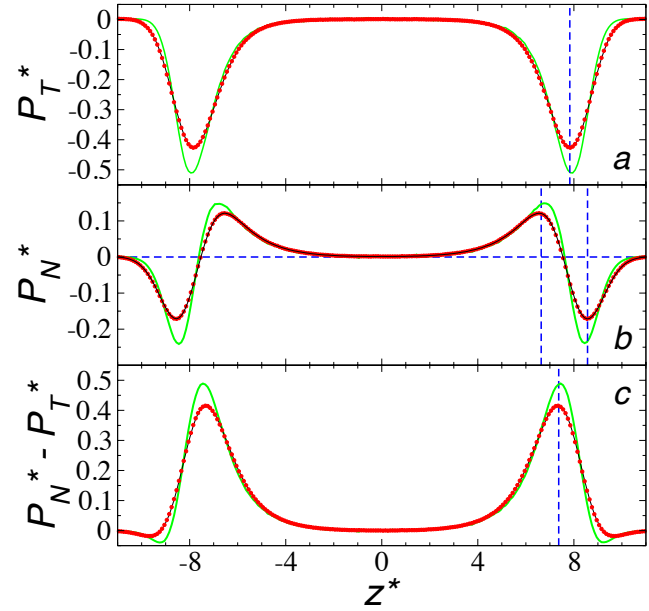

**Figure 5.** Average tangential (a) and normal (b) pressure profiles, and the differences between them (c). The systems are composed of Lennard-Jones atoms at a reduced temperature of 0.72 and reduced thicknesses of ≈ 16.8. Green, black and red lines correspond to simulated systems using reduced lengths of the simulation box in the tangential direction of 16, 140.8 and 150.4, respectively. Blue lines represent the positions of peaks in the profiles corresponding to the simulated system, using a reduced length of the simulation box in the tangential direction of 150.4.

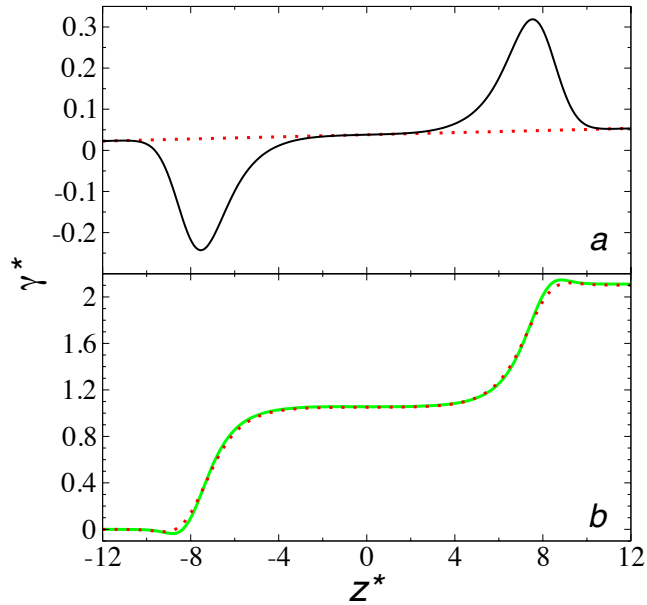

**Figure 6.** (a) Integration of the average normal pressure profile as a function of the reduced position in the simulation cell (black line). The systems are composed of Lennard-Jones atoms at a reduced temperature of 0.72, using a reduced length of the simulation box in the tangential direction of 150.4, and reduced thicknesses of 16.8. The red line corresponds to the integration of a profile with a constant normal pressure along the bulk phases and interfaces. (b) Surface tension profiles as a function of the reduced position in the simulation cell. Green and red lines correspond to simulated systems using reduced lengths of the simulation box in the tangential direction of 16 and 150.4, respectively.

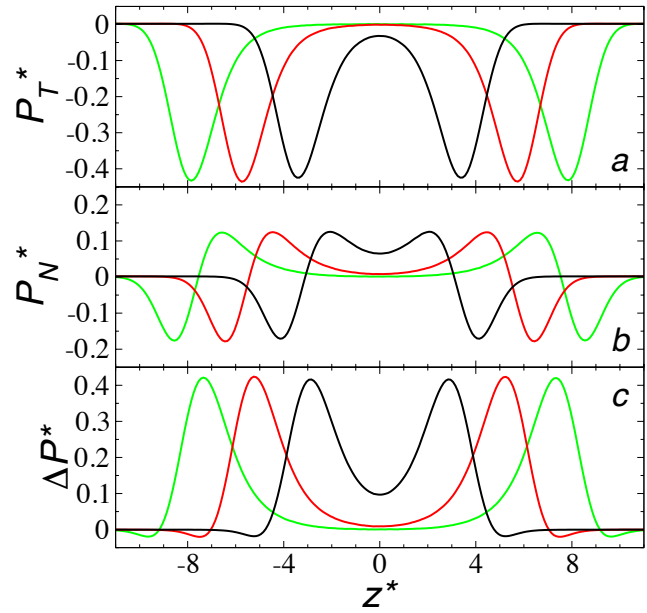

**Figure 7.** Average tangential (a) and normal (b) pressure profiles, and the differences between these (c). The systems are composed of Lennard-Jones atoms at a reduced temperature of 0.72. Profiles were calculated using a reduced length of the simulation box in the tangential direction of 120. Black, red and green lines correspond to simulated systems with lengths of the layer of 8.0, 12.6 and 16.8, respectively.

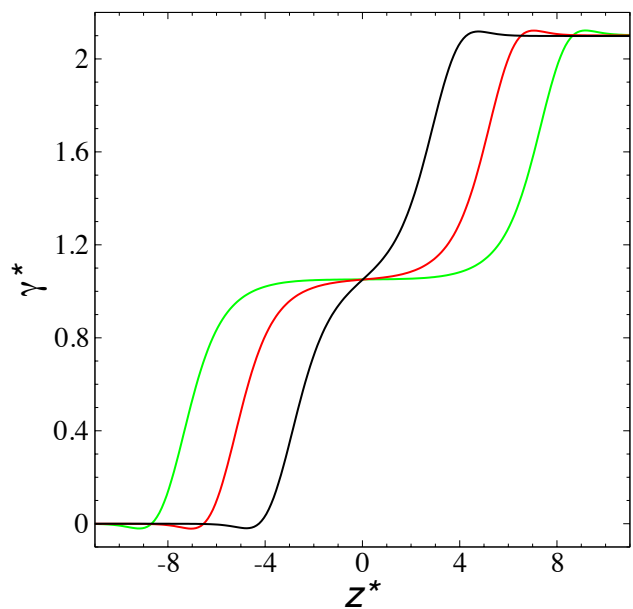

**Figure 8.** Surface tension profiles as a function of the reduced position in the simulation cell. The systems are composed of Lennard-Jones atoms at a reduced temperature of 0.72. Profiles were calculated using a reduced length of the simulation box in the tangential direction of 120. Black, red and green lines correspond to simulated systems with lengths of the layers of 8.0, 12.6 and 16.8, respectively.

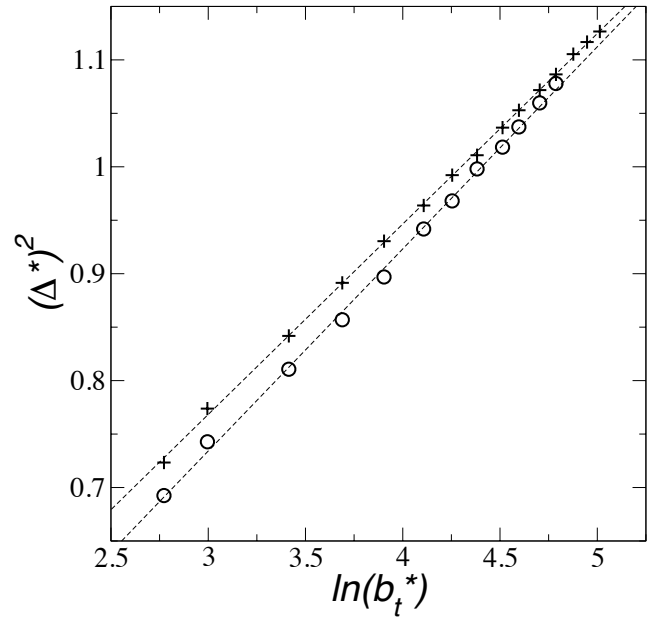

**Figure 9.** Square of the reduced interfacial thickness as a function of the length of the simulation cell in the tangential direction. The systems are composed of Lennard-Jones atoms at a reduced temperature of 0.72. Circles and crosses correspond to layers with reduced thicknesses of 8 and 16.8, respectively. Discontinuous lines correspond to linear regressions.

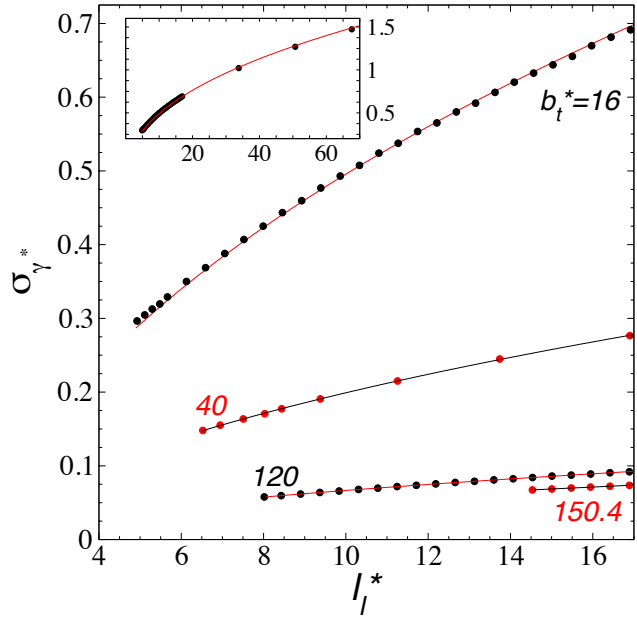

**Figure 10.** Standard deviation of the reduced surface tension as a function of the reduced length of the liquid layer of Lennard-Jones atoms. The reduced temperature of the systems is 0.72. The four sets of data represent four simulation cells with reduced lengths in the tangential direction of 16, 40, 120 and 150.4. The lines for the first three systems represent the best fits to Equation 4, while the line for the largest simulation cell size represents a linear regression. The inner graph shows the extended data for the smallest simulation cell size studied (16) and the best fit to Equation 4 at lower values. The axis labels are the same for both graphs.

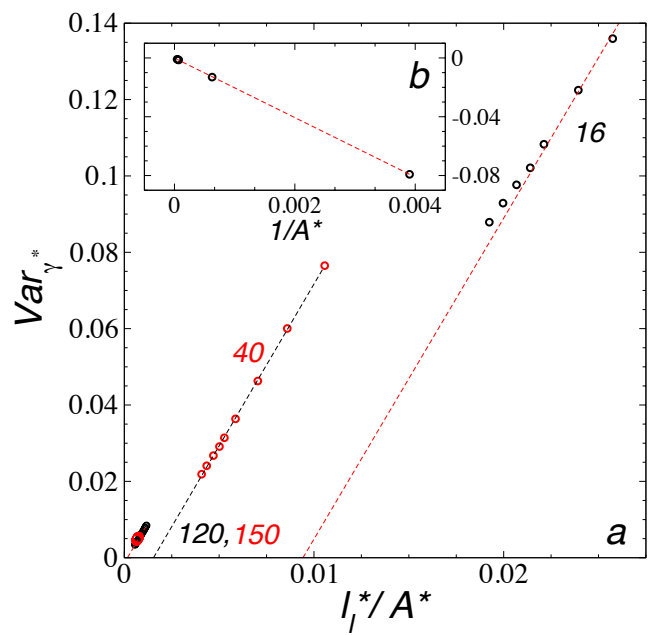

**Figure 11.** a) Variance of the reduced surface tension as a function of the ratio of the reduced length of the layer to the reduced interfacial area of the simulation cell as proposed by Longford et al.[1] Results are for reduced lengths of the simulation cell in the tangential direction of 16, 40, 120 and 150.4. The reduced temperature of the systems is 0.72. b) Intercepts of the variance of the reduced surface tension (a) with the *y*-axis as a function of the interfacial area of the simulation cell. Discontinuous lines represent the best linear regression of each set of points.

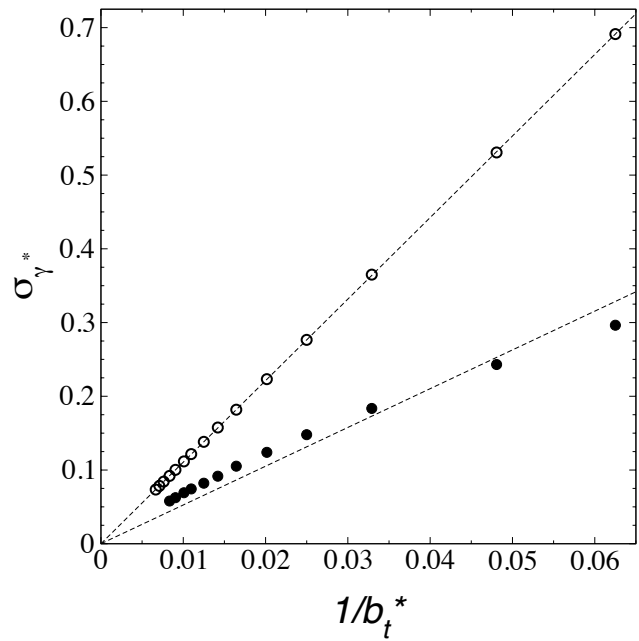

**Figure 12.** Standard deviation of the reduced surface tension as a function of the inverse of the length of the simulation cell in the tangential direction. The systems are composed of Lennard-Jones atoms at a reduced temperature of 0.72. Closed and open circles represent data for thin layers with thicknesses of 8.0 and 16.8, respectively. The discontinuous lines represent the best fits to an inversely asymptotic function (Equation 6).

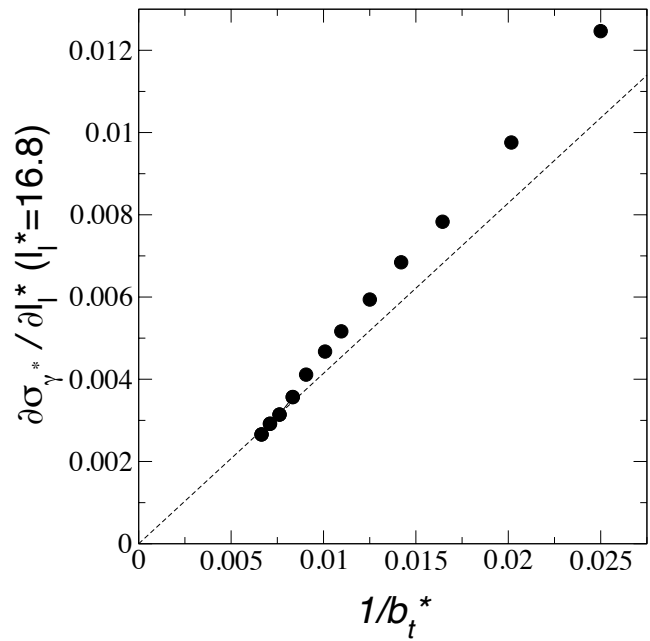

**Figure 13**. Change in the standard deviation of the reduced surface tension as a function of the inverse of the length of the simulation cell in the tangential direction. The systems are composed of Lennard-Jones atoms at a reduced temperature of 0.72 and reduced thicknesses of ≈ 16.8. The discontinuous line represents the best fit to an inversely asymptotic function (Equation 7) using the 4 largest simulation cells in the tangential direction.